\documentclass[preprint,showpacs, showkeys, preprintnumbers,amsmath,amssymb]{revtex4}

\usepackage{graphicx}
\usepackage{dcolumn}
\usepackage{bm}

\begin{document}
\title{Geometric Representation of Interacting Non-Relativistic Open Strings using Extended Objects}
\author{P. J. Arias$^{1}$ }
\email{pio.arias@ciens.ucv.ve}
\author{N. Bol\'{\i}var$^{1}$}
\email{nelson.bolivar@ciens.ucv.ve}
\author{E. Fuenmayor$^{1}$}
\email{ernesto.fuenmayor@ciens.ucv.ve}
\author{L. Leal$^{1,2}$}
\email{lleal@fisica.ciens.ucv.ve}

\affiliation{1. Centro de F\'{\i}sica Te\'{o}rica y
Computacional, Facultad de Ciencias, Universidad Central de
Venezuela, AP 47270, Caracas 1041-A, Venezuela. \\
 2. Departamento
de F\'{\i}sica, Universidad Sim\'on Bol\'{\i}var,\\ Aptdo. 89000,
Caracas 1080-A, Venezuela.\\ }

\date{\today}

\begin{abstract}
 Non-relativistic charged open strings coupled with
Abelian gauge fields are quantized in a geometric representation
that generalizes the Loop Representation. The model consists of
open-strings interacting through a Kalb-Ramond field in four
dimensions. The geometric representation proposed uses lines and
surfaces that can be interpreted as an extension of the picture of
Faraday's lines of classical electromagnetism. This representation
results to be consistent, provided the coupling constant (the
``charge'' of the string) is quantized. The Schr\"odinger equation
in this representation is also presented.
\end{abstract}

\maketitle

\section{Introduction}
 In this paper we consider theories of non-relativistic strings interacting
 with an Abelian Kalb-Ramond field \cite{KR}.
 The canonical quantization of the theory is made within the Dirac
scheme for dealing with constrained theories, and a detailed
discussion of issues as gauge invariance and the determination of
the true degrees of freedom is presented. Then, we quantize the
model in a representation that uses extended geometrical objets
(paths and surfaces) that generalizes the usual Loop
Representation ($LR$) \cite{GT}.  Special attention is devoted to
the case of open strings, whose geometric representation requires
the inclusion of open paths together with open surfaces in order
to maintain gauge invariance, as we shall discuss. A "surface
representation" was considered years ago to study the free-field
case \cite{Pio, Pio2}, but it has to be adapted to include the
particularities that the coupling with the string requires. This
study can be seen as a generalization of the theory of charged
non-relativistic point particles in electromagnetic interaction,
quantized within the $LR$ , for which it was found  that electric
charge must be quantized in order to the $LR$ formulation be
consistent \cite{Corichi1, Corichi2,E, EP, EA, ED}. As we shall
see, both for the open and closed strings models coupled with the
Kalb-Ramond field the ``charge'' of the string must be also
quantized, if the geometric representation adapted to the model is
going to be consistent. This result does not seem to be exclusive
of the  non-relativistic string case, but could be also reproduced
for the relativistic one, since it is a consequence of the
realization, in the "surface representation", of the "generalized
Gauss constraint", which is the same in both  the relativistic and
non-relativistic cases.

In the next section we
consider the model of a closed string in self-interaction by means
of an Abelian Kalb-Ramond field \cite{KR,EP, EA, ED}; the  more interesting
open string model is treated in an independent section. Final remarks
and some discussions are made at the end of the paper.

\section{Non-relativistic self-interacting ``charged'' closed string: Surface representation.}\label{sec2}

Our discussion starts reviewing the case of a closed
non-relativistic string in self-interaction \cite{EP, ED}, which is described by an
action that generalizes the theory of the self-interacting point
particle \cite{E}
\begin{eqnarray}
 S=\frac{1}{12{g^2}}\int
H^{\mu\nu\lambda}H_{\mu\nu\lambda}d^{4}x + \frac{\alpha}{2}\int dt
\int d\sigma \left[(\dot{z}^i)^2-(z'^i)^2) \right] + \frac{1}{2}
\int d^4x \textit{J}^{\mu\nu}B_{\mu\nu},\label{S}
\end{eqnarray}
where the Kalb-Ramond antisymmetric potential and field strength,
$B_{\mu\nu}$ and $H_{\mu\nu\lambda}$, respectively, are related by
$H_{\mu\nu\lambda}=3\partial_{[\mu}B_{\nu\lambda]}=\partial_{\mu}
B_{\nu\lambda}+\partial_{\lambda} B_{\mu\nu}+\partial_{\nu}
B_{\lambda\nu}$. The field $B_{\mu\nu}$ mediates the self-interaction of
 the closed string \cite{KR}, so the Maxwell type
term corresponds to its dynamical term. We have also a
contribution corresponding to the free non-relativistic closed
string, whose world sheet spatial coordinates $z^{i}(t,\sigma)$
are given in terms of the time $t$ and the parameter $\sigma$
along the string. The string tension $\alpha$ has units of
$mass^2$ and $g$ is a parameter with units of $mass$. The
string-field interaction term is given by means of the current
\begin{eqnarray}
\textit{J}^{\mu\nu}(\vec{x},t)&=& \phi \int d\tau \int d\sigma
\left[\dot{z}^\mu z'^\nu -\dot{z}^\nu
z'^\mu\right]\delta^{(4)}(x-z)\nonumber \\
&\equiv& \int d\sigma
{\cal{J}}^{\mu\nu}(t,\sigma)\delta^{(3)}(\vec{x}-\vec{z}(t,\sigma)).\label{jmunu}
\end{eqnarray}
Here $\phi$ is a dimensionless coupling constant (analog to the
charge in the case of particles), and we indicate with dots and
primes partial derivation with respect to the parameters $\tau$
and $\sigma$, respectively. We take $\dot{z}^0=1$ and $z'^0=0$.
The interaction term can be written as
\begin{equation}\label{2.3}
S_{int}=\frac{1}{2} \int dt \int d\sigma
{\cal{J}}^{\mu\nu}(t,\sigma)B_{\mu\nu}(\vec{z}(t,\sigma)).
\end{equation}

The action (\ref{S}) is invariant under the gauge transformations
$\delta B_{\mu\nu}=2\partial_{[\mu}\lambda_{\nu]}=
\partial_{\mu}\lambda_{\nu}-\partial_{\nu}\lambda_{\mu}$, provided the string is closed.

We are interested in the Dirac quantization scheme of the theory.
In this sense we observe that $B_{0i}$ is a non-dynamical
variable, so we define the conjugate momenta associated to the
fields, $B_{ij}$, and string variables, $z^i$, as
\begin{equation}\label{2.6}
\Pi^{ij}=\frac{1}{2g^2}\left(\dot{B}_{ij}+ \partial_j
B_{0i}-\partial_i B_{0j}\right), \quad \quad P_{i}=\alpha
\dot{z}^i +\phi B_{ij}z'^{j},
\end{equation}
and obtain the Hamiltonian performing a Legendre transformation in
the dynamical variables $B_{ij}$ and $z^{i}$,
\begin{eqnarray}\label{2.7}
H=\int d^3x \left[{g}^2
\Pi^{ij}\Pi^{ij}+\frac{1}{12g^2}H_{ijk}H_{ijk}\right]&+&\int
d\sigma \frac{\alpha}{2}\left[\frac{1}{\alpha^2}\left(P_{i}-\phi
B_{ij}(z)z'^{j}\right)^{2}+(z'^{i})^2 \right]+ \nonumber \\
&+&\int d^3x B_{0i}\chi^i ,
\end{eqnarray}
where the role of $B_{0i}$ becomes clear as Lagrange multipliers
enforcing the constraints
\begin{equation}\label{2.8}
\chi^{i}(x) \equiv - \rho^{i}(x) - 2\partial_j \Pi^{ji}(x))=0.
\end{equation}
Here, $\rho^{i}(x)\equiv \phi \int d\sigma
z'^i\delta^{(3)}(\vec{x}-\vec{z})$ ($=J^{0i}(x)$) is the ``charge
density'' of the string. The preservation of the above constraints
can be done using the canonical Poisson algebra of the
fields involved. The non-vanishing Poisson brackets are given by
\begin{eqnarray}\label{2.10}
\left\{z^i(\sigma),P_j(\sigma')\right\}=\delta^i_j\delta(\sigma
-\sigma'),
\end{eqnarray}
\begin{eqnarray}\label{2.11}
\left\{B_{ij}(\vec{x}),\Pi^{kl}(\vec{y})\right\}=\frac12\left(\delta_{i}^{k}\delta_{j}^{l}
-\delta_{i}^{l} \delta_{j}^{k}\right)
\delta^{(3)}(\vec{x}-\vec{y}).
\end{eqnarray}
The preservation of the constraints does not produce new ones,
and the $B_{0i}$ remain undetermined. This tells us that the
constraints are first class (as can be directly verified by calculating their
Poisson brackets) and generate time independent gauge
transformations.

The basic observables, in the sense of Dirac, that can be
constructed from the canonical variables, are the generalized
electric and magnetic fields
\begin{eqnarray}\label{2.12}
\Pi^{ij}&=&\frac{1}{2g^2}H_{0ij}\equiv \frac{1}{2g^2}E^{ij},
\label{genelectric} \\
\textbf{B}&\equiv& \frac{1}{3!} \epsilon^{ijk}H_{ijk},
\label{genmagnetic}
\end{eqnarray}
the position $z^{i}(\sigma)$, and the covariant momentum of the
string
\begin{equation}\label{2.13}
P_{i}-\phi B_{ij}(z)z'^{j}.
\end{equation}
All the physical observables of the theory are built in terms of
these gauge invariant quantities, as can be verified. For
instance, the Hamiltonian, given in equation (\ref{2.7}), fulfils
this requirement.

To quantize, we promote the canonical variables
to operators that obey a conmutator algebra that results from the
replacement $\{ \quad , \quad \} \to -i[ \quad , \quad ]$, as usual. These
operators have to be realized in a Hilbert space whose physical
states $|\Psi\rangle_{Phys}$ are in the kernel of the constraint
$\chi^{i}$
\begin{equation}
\chi^{i}|\Psi\rangle_{Phys}=0.\label{phys}
\end{equation}

Now, in order to solve relation (\ref{phys}), a geometric
representation adapted to the present model is introduced. This
representation, based on extended objects, will be a ``surface
representation'' related with the $LR$ formulated by Gambini and
Tr\'{\i}as \cite{GT}, and with an early geometrical formulation of
the pure Kalb-Ramond field based on closed surfaces \cite{Pio,
Pio2}. Consider the space of piecewise smooth oriented surfaces
in $R^{3}$. A typical element of this space,
let us say $\Sigma$, will be the union of several surfaces, with
some of them being closed. In the space of smooth oriented
surfaces $\Sigma$ we define equivalence classes of surfaces that
share the same ``form factor'' $T^{ij}(x,\Sigma)=\int
d\Sigma^{ij}_y\,\delta^{(3)}(\vec{x}-\vec{y})$, where
$d\Sigma^{ij}_y=(\frac{\partial y^i }{\partial s}\frac{\partial
y^j}{\partial r} - \frac{\partial y^i}{\partial r}\frac{\partial
y^j}{\partial s})dsdr$ is the surface element and $s$, $r$ are the
parametrization variables. All the features of the ``open surfaces
space'', are generalizations of aspects already present in the
Abelian path space \cite{GT, EP, LO, C, EA, ED}.

Our Hilbert space is composed by functionals $\Psi(\Sigma)$ depending
on equivalence classes $\Sigma$. We need to introduce the surface
derivative $\delta_{ij}(x)$ defined by,
\begin{equation}\label{2.17}
\Psi (\delta\Sigma\cdot\Sigma)-\Psi
(\Sigma)=\sigma^{ij}\delta_{ij}(x)\Psi (\Sigma)
\end{equation}
that measures the response of $\Psi(\Sigma)$ when an element of
surface whose infinitesimal area $\sigma_{ij}=u^iv^j-v^ju^i$,
generated by the infinitesimal vectors $\vec{u}$ and $\vec{v}$, is
attached to $\Sigma$ at the point $x$  \cite{Pio, Pio2, EP, EA,
ED}.

It can be seen that the fundamental commutator associated to
relation (\ref{2.11}) can be realized on surface-dependent
functionals if one sets
\begin{equation}\label{2.20}
\hat{\Pi}^{ij}(\vec{x})\longrightarrow
\frac{1}{2}T^{ij}(\vec{x},\Sigma),
\end{equation}
\begin{equation}\label{2.21}
\hat{B}_{ij}(\vec{x})\longrightarrow 2i\delta_{ij}(\vec{x}),
\end{equation}
since the surface-derivative of the form factor is given by
\begin{equation}\label{2.22}
\delta_{ij}(\vec{x}) T^{kl}(\vec{y},\Sigma)=\frac
12\left(\delta_{i}^{k}\delta_{j}^{l} -\delta_{i}^{l}
\delta_{j}^{k}\right)\delta^{(3)}(\vec{x}-\vec{y}).
\end{equation}
In this sense the states of the interacting theory can be taken as
functionals $\Psi[\Sigma,z(\sigma)]$, where the field is
represented by the surface $\Sigma$ and matter by means of the
coordinates of the string world sheet. On the other hand, the
operators associated to the string can be realized
onto these functionals $\Psi[\Sigma,z(\sigma)]$ as follows

\begin{equation}\label{2.23}
\hat{z}^i(\sigma) \longrightarrow z^i(\sigma), \quad
\hat{P}^i(\sigma)\longrightarrow -i \frac{\delta} {\delta
z^i(\sigma)} .
\end{equation}
The operators of the theory are then realized in a representation that is the
tensor product of the "open-surface" representation, for the the field operators
and a "shape" representation for the string operators.
Of all of these functionals we choose, as we stated, those that
belong to the kernel of the generalized Gauss constraint
(\ref{2.8}), now written as
\begin{eqnarray}\label{2.24}
\left(-\rho^{i}(\vec{x}) - 2\partial_j\Pi^{ji}(\vec{x})\right)
\Psi[\Sigma,z(\sigma)]&\approx& 0\nonumber\\
\left(-\phi\int_{string} d\sigma z'^i\delta^{(3)}(\vec{x}-\vec{z})
+\int_{\partial\Sigma}d\sigma z'^i
\delta^{(3)}(\vec{x}-\vec{z})\right)\Psi[\Sigma ,z(\sigma)]
&\approx& 0.
\end{eqnarray}
In the last equation we have used that $\partial_j
T^{ji}(\vec{x},\Sigma)=-T^i(\vec{x},\partial\Sigma)
=-\int_{\partial\Sigma} dz^i \delta^{(3)}(\vec{x}-\vec{z})$, with
$\partial \Sigma$ being the boundary of the surface.

If the oriented surface is such that its boundary coincides with
the orientation of the string, the constraint reduces to
$\left(\phi-1\right)\,\int_{string}d\sigma z'^j
\delta^{(3)}(\vec{x}-\vec{z})=0$, and it is satisfied in general
for $\phi=1$. We say in this case that the surface ``emanates'' or
``starts'' from the string, in analogy with the theory of
self-interacting non-relativistic particles coupled through a
Maxwell field \cite{E, EA}. It could happen instead, that the
boundary of the surface and the string have opposite orientations;
in that case the constraint would be satisfied if $\phi =-1$, and
we say that the surface ``enters'' or ``arrives'' at the string
position. There exist also the possibility that the surface could be
composed by several layers ($n$ of them) that start (or end) at
the string. Equation (\ref{2.24}) becomes
$\left(\phi-n\right)\,\int_{string}d\sigma z'^j
\delta^{(3)}(\vec{x}-\vec{z})=0$, and in this case the coupling
constant (``charge'' of the string) must obey $\phi =n$ (the sign
of $n$ depends on the fact that the surfaces  may ``emanate'' from
or ``arrive'' to the source). This is what we call a
representation of ``Faraday's surfaces'' for the
string-Kalb-Ramond system, in analogy with the particle-Maxwell
case. Finally, it should be remarked that when $\phi =n$, the
surface may consist of the $n$ layers attached to the string, plus
an arbitrary number of closed surfaces, since the latter do not
contribute to the boundary of the surface that define the
equivalence class $\Sigma$.

\section{Open strings: Surface$+$Path representation.}\label{sec3}

With this insight we move to the main subject  of this paper, the model of open
non-relativistic  self-interacting strings. Our
starting point will be the action,
\begin{eqnarray}\label{2.25}
 \mathcal{S}&=&\int dt \int d\sigma \frac{\alpha}{2}\lbrack
 (\dot{z}^{i})^{2}-(z'^{i})^{2}\rbrack+\nonumber \\
  & & \qquad \qquad +\int d^{4}x \left(\frac{1}{12g^{2}}H^{\mu\nu\lambda}H_{\mu\nu\lambda}-
  \frac{m^{2}}{4}a^{\mu\nu}a_{\mu\nu}+
  \frac{1}{2}J^{\mu\nu}B_{\mu\nu}+J^{\mu}A_{\mu}\right),\label{Sa}
\end{eqnarray}
where we have defined the $2$-form
$a_{\mu\nu}=B_{\mu\nu}+F_{\mu\nu}$ with
$F_{\mu\nu}=\partial_{\mu}A_{\nu}-\partial_{\nu}A_{\mu}$ as in the
St\"{u}ckelberg gauge invariant version of the Proca model
\cite{E, EA}. The vector field $A_{\mu}$ has dimensions of $mass$,
as in Maxwell theory, and mediates the interaction between the
ends of the string. The Kalb-Ramond antisymmetric potential and
field strength are related as before. The parameter $m$ is
dimensionless.

The gauge field terms in (\ref{Sa}) resemble the model for open
strings proposed by Kalb and Ramond \cite{KR}. The action is
invariant under the simultaneous gauge transformations,
\begin{eqnarray}\label{2.26}
B_{\mu\nu}\; \longrightarrow \; B_{\mu\nu}+\partial_{\mu}
\Lambda_{\nu} -
\partial_{\nu} \Lambda _{\mu}\, ,\qquad
A_{\mu}\; \longrightarrow\;
A_{\mu}-\Lambda_{\mu}+\partial_{\mu}\Lambda\,
\end{eqnarray}
only if the currents associated to the matter source (the body and the
ending points of the string) satisfy the relation
\begin{eqnarray}\label{2.27}
\partial_{\mu}J^{\mu\nu}+J^{\nu}=0,
\end{eqnarray}
which emerges as a consequence of enforcing the equations of
motion to be  gauge invariant. There is still a residual gauge
invariance in the gauge parameters
($\delta\Lambda_{\mu}=\partial_{\mu}\lambda$,
$\delta\Lambda=\lambda$) that can be removed if we state that the
gauge function $\Lambda_{\mu}$ is transverse (see the Appendix for
further comments) .

The source condition (\ref{2.27}) implies that the current
associated to the endpoints is conserved ($\partial_{\nu}J^{\nu}=0$),
even if the string-current $J^{\mu\nu}$ is not. This is a consequence
of the form of the string interaction term, which has two types of currents
associated to matter: one that is associated to the ``body'' and
the other to the end points of the string. This fact
 permits the treatment of open-strings
without losing gauge invariance. The source that couples with the
Kalb-Ramond field is given by equation (\ref{jmunu}) as before, while the source that
couples with the vector field is
\begin{eqnarray}
\textit{J}^{\mu}(\vec{x},t)&=& e \int
dz^{\mu}\delta^{(4)}(x-z){\mid}^{z_f}_{z_i},
\nonumber \\
&\equiv& \int d\sigma
{\cal{J}}^{\mu}(t,\sigma)\delta^{(3)}(\vec{x}-\vec{z}(t,\sigma)),
\end{eqnarray}
where we defined ${\cal{J}}^{\mu}(t,\sigma)\equiv e
\dot{z}^{\mu}(t,\sigma)(\delta(\sigma-\sigma_f)-\delta(\sigma-\sigma_i))$
(with $\dot{z}^0=1$), and the subscripts $i$ and $f$ identify the
initial and final points of the string. The vector source can be
thought as two opposite charges attached to the ends of the open
string. It can be checked that the sources satisfy the constraint
(\ref{2.27}), provided that $\phi=-e$. In the case of more than
one open string the generalization is straightforward.

The equations of motion that arise from (\ref{2.25}) are
\begin{eqnarray}
\frac{1}{g^2}\partial_{\lambda}H^{\lambda\mu\nu}+m^{2}a^{\mu\nu}-J^{\mu\nu}&=&
0,\label{a} \\
m^{2}\partial_{\lambda}a^{\lambda\mu}+J^{\mu}&=&0,\label{b}\\
\alpha(z''^i-\ddot{z}^i)+\frac{1}{2}H_{i\mu\nu}(z){\cal{J}}^{\mu\nu}(t,\sigma)
+F_{i\mu}(z){\cal{J}}^{\mu}(t,\sigma)&=&0
\end{eqnarray}
and they guarantee that (\ref{2.27}) is satisfied, in order to be
gauge invariant under (\ref{2.26}). From (\ref{a}) and
(\ref{b}) we obtain that the excitations are massive
\begin{equation}
(\Box +m^2g^2)a^{\mu\nu}=g^2{\textit{J}}^{\mu\nu}
-\frac{1}{m^2}(\partial^{\mu}\textit{J}^{\nu}-\partial^{\nu}\textit{J}^{\mu}).
\label{ecmasiva}
\end{equation}

The field $A_{\mu}$ can be eliminated in virtue of the gauge
invariance, in this sense it is said that the Maxwell photon is
``absorbed'' to obtain a massive pseudovector \cite{KR}.

Now, we implement with the Dirac quantization procedure. We take
$A_{i}$, $B_{ij}$ and $z^{i}$ as dynamical variables and
$\Pi^{i}$, $\Pi^{ij}$ and $P_{i}$ as their canonical conjugate
momenta, respectively.  After a Legendre
transformation in the dynamical variables the Hamiltonian results to be
\begin{eqnarray}\label{2.29}
H&=&\int d^3x\left[
g^{2}\Pi^{ij}\Pi^{ij}+\frac{1}{2m^{2}}\Pi^{i}\Pi^{i}+\frac{1}{12g^{2}}H_{ijk}H_{ijk}
+\frac{m^{2}}{4}a_{ij}a_{ij} + A_{0}\Theta + B_{0i}\Theta^{i}\right]+\nonumber \\
&& + \int d\sigma
\frac{\alpha}{2}\left\{\frac{1}{\alpha^2}\left[P_{i}-\phi
B_{ij}(z)z'^{j}-eA_i(z)\left(\delta(\sigma-\sigma_f)-\delta(\sigma-\sigma_i)\right)\right]^{2}+(z'^{i})^2
\right\},
\end{eqnarray}
where, in analogy with the closed-string model, the fields
variables $A_{0}$ and $B_{i0}$ appear in $H$ as Legendre
multipliers enforcing the constraints
\begin{eqnarray}\label{thetas}
\Theta&\equiv&-\partial_{j}\Pi^{j}-J^{0}\approx 0,\nonumber \\
\Theta^{i}&\equiv&-2\partial_{j}\Pi^{ji}-\Pi^{i}-J^{0i} \approx 0.
\end{eqnarray}
It can be seen that $\partial_{i}\Theta^{i}=\Theta$ when
$\phi=-e$, hence, the constraints form a reducible set.

The usual Poisson algebra between the canonical conjugate variable
is now defined. The non-vanishing brackets are
\begin{eqnarray}
\{ B_{ij}(\vec{x}), \Pi^{kl}(\vec{y}) \}&=&
\frac{1}{2}(\delta^{k}_{i}\delta^{l}_{j}-\delta^{l}_{i}\delta^{k}_{j})\delta^{(3)}(\vec{x}-\vec{y}), \label{Bpi}\\
\{ A_{i}(\vec{x}), \Pi^{k}(\vec{y}) \} &=&
\delta^{k}_{i}\delta^{(3)}(\vec{x}-\vec{y}), \label{Api}\\
\{z^i(\sigma),P_k(\sigma')\}&=&\delta^{i}_{k}\delta(\sigma-\sigma').
\end{eqnarray}

When the constraints are preserved no more constraints emerge and
the Lagrange multipliers remain undetermined. So $\Theta$ and
$\Theta^i$ turns out to be first class constraints that generates time
independent gauge transformations on phase space. The generator of
the infinitesimal gauge transformations is $G=\int d^3x
(\Lambda\Theta+\Lambda_i\Theta^i)$ and its effect on the dynamical
fields is $\delta F=\{F,G\}$. The dynamical variables transform as follows
\begin{eqnarray}
\delta A_i=\partial_i\Lambda - \Lambda_i \quad &,& \quad \delta
\Pi^i=0,
\label{1}\\
\delta B_{ij}=\partial_i\Lambda_j - \partial_j\Lambda_i \quad &,&
\quad \delta \Pi^{ij}=0,\label{2} \\
\delta z^i=0 \quad &,& \quad \delta
P_i=(\partial_i\Lambda_k(z)-\partial_k\Lambda_i(z)){\cal{J}}^{0k}+
(\partial_i\Lambda(z)+\frac{\phi}{e}\Lambda_i(z)){\cal{J}}^{0},
\label{3}
\end{eqnarray}
where it should be understood that $F(z)=\int d^3x
F(x)\delta^{3}(\vec{x}-\vec{z})$. Using these transformations it
can be seen that $H$ is gauge invariant as it is expected.

To count the number of degrees of freedom we note that we had
initially $9$ dynamical fields ($A_i$, $B_{ij}$ and $z^i$) ,
subject to $3$ first class constraints ($\Theta$ and the
transverse part of $\Theta_i$), or $4$ minus the reducibility
condition, plus $3$ gauge conditions. This leave us with
$2\times9-2\times3=12$ coordinates and conjugate momenta,
corresponding to $6$ degrees of freedom, $3$ for the string
coordinates ($z^i$) and $3$  for the massive pseudo vector field
($B_{ij}$). A further discussion of this subject and the gauge
dependence of the model is presented in the Appendix at the end.

To quantize, as usual, the canonical variables are promoted to
operators obeying the commutators that result from the replacement
$\{\;,\;\}\;\longrightarrow\;-i[\;,\;]$. These operators have to
be realized in a Hilbert space of physical states
$|\Psi\rangle_{Phys}$, that obey the constraints (\ref{thetas}) ($
\Theta^{A}|\Psi\rangle_{Phys}=0$). The observables of the theory
are constructed with gauge invariant objects.

At this point, we adapt a geometrical representation to the theory
in terms of extended objects as it was done in the case of
self interacting closed-strings. Now we take, besides the
open surfaces associated to the Kalb-Ramond field, open paths $\gamma$ \cite{GT}
associated to the fields $A_{i}$,
that mediate the interaction between the endpoints of the
string. We prescribe,
\begin{eqnarray}\label{2.34}
\hat{B}_{ij}(\vec{x})\longrightarrow 2i\delta_{ij}(\vec{x}) &,&
\hat{\Pi}^{ij}(\vec{x})\longrightarrow
\frac{1}{2}T^{ij}(\vec{x},\Sigma)\\
\hat{A}_{i}(\vec{x})\longrightarrow i\delta_{i}(\vec{x}) &,&
\hat{\Pi}^{i}(\vec{x})\longrightarrow T^{i}(\vec{x},\gamma)
\end{eqnarray}
where $T^{i}(\vec{x},\gamma)$ and and $T^{ij}(\vec{x},\Sigma)$ are
the form factors that describes the open paths $\gamma$ and open
surfaces $\Sigma$, respectively \cite{GT, EP, E}. We can see,
using $\partial_{j}T^{ji}(\vec{x},\Sigma)= -T^{i}(\vec{x},\partial
\Sigma)$ and $\delta_{ij}(\vec{x})T^{lk}(\vec{y},\Sigma)=
\frac{1}{2}(\delta^{l}_{i}\delta^{k}_{j}
-\delta^{k}_{i}\delta^{l}_{j})\delta^{(3)}(\vec{x}-\vec{y})$, that
the fundamental commutators associated to equations (\ref{Bpi})
and (\ref{Api}) can be realized when they act over functionals
depending on both surfaces and paths. Also, the operators
associated to the string can be realized in the same ``shape"
representation, as before. Henceforth, the states of the
interacting theory can be taken as functionals $\Psi[\Sigma,
\gamma, z(\sigma)]$ depending on geometric objects (open surfaces
and  paths) and the string coordinates $z^i(\sigma)$. Among these
functionals, we must pick out those that belong to the kernel of
the constraints (\ref{thetas}), i.e., the physical space. In this
representation they can be expressed as
\begin{eqnarray}
\left(-\partial_i T^i(\vec{x},\gamma)-e\int d\sigma
(\delta(\sigma-\sigma_f)-\delta(\sigma-\sigma_i))
\delta^{(3)}(\vec{x}-\vec{z})\right)\Psi[\Sigma,
\gamma, z^i(\sigma)]&\approx& 0 \label{reptheta}\\
\left(T^{i}(\vec{x},\partial\Sigma) - T^{i}(\vec{x},\gamma) +
\phi\int_{\gamma_s} d z^{i} \delta^{(3)}(\vec{x}-\vec{z})
\right)\Psi[\Sigma, \gamma, z^i(\sigma)]&\approx&
0,\label{repthetai}
\end{eqnarray}
with $\partial \Sigma$ being the boundary of $\Sigma$, and $\gamma_s$ is the path representing the string. The
meaning of $\Sigma$ and $\gamma$ can be understood as follows.
Given an open path $\gamma$ starting at $\vec{z}_i$ and ending at
$\vec{z}_f$ it happens that $\partial_i
T^i(\vec{x},\gamma)=-(\delta^3(\vec{x}-\vec{z}_f)-\delta^3(\vec{x}-\vec{z}_i))$,
so (\ref{reptheta}) is solved if $\gamma$ is attached
to the endpoints of the string. If $e$ is positive  the path should start at $\vec{z}_i$ and end at $\vec{z}_f$,
which is consistent with the geometrical representation of
non-relativistic particles in electromagnetic interaction
\cite{EP}, and the constraint gives $e=1$. In a more general sense
$\gamma$ may consist of a bundle of open paths  starting
at $\vec{z}_i$ and ending at $\vec{z}_f$. In this case the charge has to be
quantized and the number of paths depends on the value of $\mid e
\mid$. It is clear that, since  we are talking about equivalence
classes of paths,  the bundle of open paths could be accompanied  by an additional number of
loops that will not affect the interpretation. Looking at
(\ref{repthetai}), we have that the relation $\partial_i\Theta^i=\Theta$ is
satisfied, as we have seen, when $\phi=-e$ (or more generally when
$\phi=q_i$ with $q_i$ the charge at the initial point of the
string). On the other hand, and similarly to the closed-string case, we
take $\Sigma$ as an open surface that has the string as part of his
border, in such a way that if $\phi$ is positive the orientation of
$\partial\Sigma$ and $\gamma_s$ coincide (when $\phi$ is negative
their orientations are opposite). With this picture we see that,
for $e$ positive, (\ref{repthetai}) is solved if
$\partial\Sigma=\gamma + \overline{\gamma}_s$, where the overline
on a path means the same path but with opposite orientation. We
could have several open surfaces with the strings and the
bundle of paths as their borders. The number of surfaces will
depend on the absolute value of $e$ (or equivalently of $\phi$).

\begin{figure}
\centering
\includegraphics[scale=0.9]{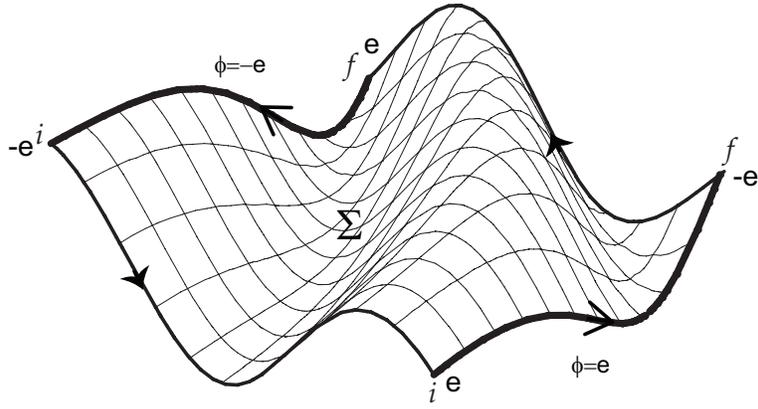}
\caption{Surface-path representation example for 2 strings
($e>0$). The bold borders correspond to the strings.} \label{fig1}
\end{figure}

\begin{figure}
\centering
\includegraphics[scale=1]{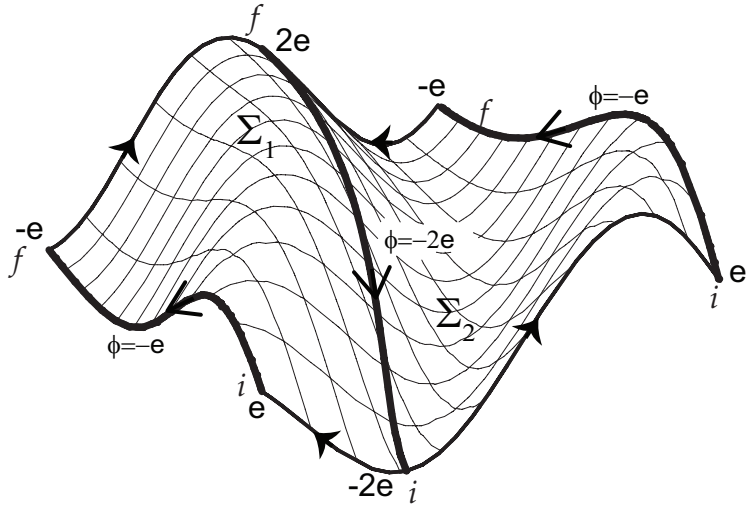}
\caption{Surface-path representation for 3 strings ($e>0$). The
bold borders correspond to the strings.} \label{fig2}
\end{figure}

In a general situation we could have a group of open strings,
enumerated with a subscript $a$ each of them, with  quantized
coupling constant $\phi_a$, in such a way that their endpoints can
be thought as  pairs of opposite charges. The ``initial point'' of
each string has charge $q_a=\phi_a$. This guarantees that each
source satisfies the constraint in order to preserve gauge
invariance. The states of the theory correspond to functionals
$\Psi[\Sigma, \gamma, z^i(\sigma)]$ that depends on open surfaces,
open paths and the coordinates of the strings, where the open
paths $\gamma$ are equivalence classes of a bundle of
$\mid\phi_a\mid$ paths attached to the endpoints of the strings
ending on the positive charge (outgoing from the negative charge).
The open surface $\Sigma$ is a set of surfaces whose borders
correspond to $\gamma$ and the strings. In figures \ref{fig1} and
\ref{fig2} we show some examples of the surfaces $\Sigma$ and
paths $\gamma$ that satisfy the gauge constraints. In figure
\ref{fig1} we show an example for 2 strings with one of them
parametrized in such a way that the initial charge is positive. In
figure \ref{fig2} we present an example for 3 strings, one of them
with charges $\pm 2e$ at its extremes. In the examples it is shown
that the orientation of the surface $\Sigma$ coincides with the
orientation of the parametrization of the string when the initial
charge is positive and it is opposite when the initial charge is
negative. This situation will be the same if we change the
orientation of the parametrization. The other part of the border
correspond to open paths that start on negative charges and end on
positive charges in a number coincident with the multiplicity of
charge $\pm e$. In \ref{fig2} the string with charge $\pm 2e$ at
its extremes (with $\phi=-2e$) belongs to part of the border of 2
open surfaces as it should. These cases are non unique, for
example in the case of 2 strings there is another configuration
with each string attached to part of the border of an open surface
with the rest of the border corresponding to an open path
connecting the extremes of the string. This configuration with 2
open surfaces is not topologically equivalent to the one presented
in figure \ref{fig1}.

Finally we can write the Schr\"odinger equation in the
path-surface representation. Taking into account $H$ in
(\ref{2.29}) acting on physical wave functionals, we write

\begin{eqnarray}
&&-i\frac{\partial}{\partial t}\Psi[\Sigma, \gamma,
z^i(\sigma)]=H\Psi[\Sigma, \gamma, z^i(\sigma)] \nonumber \\
&&\qquad \qquad=\Bigg\{\int d^3x
\left[\frac{1}{2g^2}\left((\frac{i}{3}\varepsilon^{ijk}\Delta_{ijk}(\vec{x}))^{2}
+
\frac{1}{2}E^{ij}E^{ij}\right)+\frac{1}{2m^2}T^{i}(\vec{x},\gamma)
T^{i}(\vec{x},\gamma)+\right. \nonumber \\
&&\qquad \qquad
-\left.\left(\frac{m^2}{4}(2\delta_{ij}(\vec{x})+\Delta_{ij}(\vec{x}))(2\delta_{ij}(\vec{x})+\Delta_{ij}(\vec{x}))
\right)\right]+ \nonumber \\
&&\qquad \qquad +\int
d\sigma\frac{\alpha}{2}\left[\frac{1}{\alpha^2}\left(
\frac{\delta}{\delta z^i(\sigma)}-
2i\phi\delta_{ij}(\vec{z})z'^i(\sigma)-ie\delta_i(\vec{z})(\delta(\sigma-\sigma_f)-
\delta(\sigma-\sigma_i))\right)^2\right.\nonumber \\
&&\qquad \qquad \qquad \qquad \qquad
+(z'^i(\sigma))^2\Bigg]\Bigg\} \Psi[\Sigma, \gamma, z^i(\sigma)].
\label{schro}
\end{eqnarray}
We see in (\ref{schro}) the free field contributions to the energy
of the system, that appear as generalized Laplacians formed with
surface and path derivatives, as well as quadratic terms
containing form factors of paths and surfaces. The derivative
terms come from the "magnetic" operators  $\textbf{B}$ and
$a_{ij}$
\begin{eqnarray}
\textbf{B}\equiv \frac{1}{3!} \epsilon^{ijk}H_{ijk}& \Rightarrow &
\textbf{B}=\frac{i}{3}\varepsilon^{ijk}\Delta_{ijk}(\vec{x}), \label{Bcont} \\
a_{ij} & \Rightarrow &
i(2\delta_{ij}\left(\vec{x})+\Delta_{ij}(\vec{x})\right),
\label{acont}
\end{eqnarray}
where we note the appearance of the closed surface derivative
$\Delta_{ijk}(\vec{x})$ defined in \cite{Pio, Pio2} and of a
combined surface-loop derivative. This last combination takes into
account that the open surface derivative alone is not gauge
invariant, since  paths are part of the boundary of surfaces, as
discussed above. The ``position'' contributions come from $E^{ij}$
($\rightarrow \frac{1}{2}T^{ij}(\vec{x},\Sigma)$) and
$T^i(\vec{x},\gamma)$. Their should be regularized due to the
appearance of Dirac delta functions products. The string
contributions to the Hamiltonian comprise, besides  terms
corresponding to the kinetic and potential energy of the free
string, the minimal coupling of the string variables with the
Kalb-Ramond and vector fields. These appear as generalized
Mandelstam derivatives, in the sense that as the functional
$z^i(\sigma)$ derivative translates (infinitesimally) the string,
the surface and path derivatives must also act in order to
maintain the surface $\Sigma$ and its borders (paths and strings)
joint together to preserve  gauge invariance.

\section{Discussion}\label{sec4}

We have studied a generalization of the $LR$ quantization of
strings interacting by means of the Kalb-Ramond field. When the
strings are closed, we saw that this representation is a ``surface
representation'' that may be set up only if the coupling constant
$\phi$ of the string (equivalent to the charge if we were dealing
with point particles) is quantized, so it takes only integer
values $n$. Hence, the theory is in a sense very similar to the
Maxwell theory of interacting particles  in the framework of the
$LR$  \cite{E, EP, ED}. There,  the closed paths of the free
Maxwell theory become open paths that start and end just where the
charged particles are. In this sense, it is a ``Faraday`s lines
representation''. It results then that both the electric flux
carried by each Faraday`s line and the electric charge are
quantized in order to maintain gauge invariance. In the present
study things are very similar.
 In both cases the
appropriate Hilbert space is made of wave functionals whose
arguments are geometric ``Faraday`s extended objects'' (now
 surfaces) emanating from or ending at the strings
(or particles) positions. The quantization of the ``charge''
of the particles or strings , i.e. the quantization of the coupling
constant, is necessary to solve the first class Gauss constraint.
In the case of
$N$ strings, carrying different ``charges'' $\phi_{a}$,
$a=1,..,N$, each string must be a source or sink of its own bundle
of  $n_{a}=\phi _{a}$ layers (these bundles may be
accompanied by closed pieces of surfaces). This geometrical
setting is possible if the couplings $\phi_{a}$ are quantized,
since each individual sheet or layer carries a unit of Kalb-Ramond
electric flux.

A step further, which is the main subject of this paper, is the
case of the open string interaction. Now, in order to keep gauge
invariance, we had to considerate separate couplings of gauge
fields to the body and the endpoints of the strings. The
corresponding geometric representation, in this case, yields the
following picture. The states of the interacting theory of open
strings can be taken as functionals $\Psi[\Sigma, \gamma,
z(\sigma)]$ depending on surfaces and paths, and functions of the
string variables $z(\sigma)$, that act as the source of the
extended geometrical objects. The body of the string "interacts"
with a surface, that depending on the orientation of the string
(i.e., the coupling constant $\phi$), ``emanates'' from or
``arrives'' to it. In turn, the endpoints of the string "interact"
via the open paths (just as in the case of electromagnetic
interaction for particles) that complete the part  of the border
of the surface that is not ``glued'' to the strings. Again, as in
the closed string interaction with the Kalb-Ramond field, the
surface may consist of $n$ layers attached to the string
(depending on the value of the coupling constant $\phi$), plus an
arbitrary number of closed surfaces, since the latter do not
contribute to the boundary of the surface. Each layer of open
surface carries a unit of Kalb-Ramond flux emerging from or
entering to a string, depending on the value of the string-charge
$\phi$, and this is totally compatible with the number of open
paths related to the endpoints of the string. This produces the
quantization of the coupling constant of the strings. We have also
presented the Schr\"odinger equation in the path-surface
representation, analyzing the different terms that appear.

Following references \cite{EP, LO, ED, Le, W, Wu, In} one could also consider the
geometric representation of open strings interacting through
topological terms, like a $BF$ term in $3+1$ dimensions. In these models the dependence
of the wave-functionals on paths (or more generally, on the
appropriate geometric objects that would enter in the representation) might be eliminated by means of an unitary
transformation \cite{LO, C, Le}. In that case one could obtain a quantum
mechanics of particles, or particles and strings (depending on the
model), subjected to long range interactions leading to
anomalous statistics \cite{Wu, In}. This and other topics shall be
the subject of future investigations.

\section{Appendix}

In this Appendix we complete the discussion about gauge
invariance of the open string model, and show how it can be
expressed in a covariant way as it is expected from the
original action (\ref{2.25}).

After obtaining $H$ in (\ref{2.29}) we saw that $A_0$ and $B_{0i}$
are lagrange multipliers associated with the first class
constraints $\Theta$ and $\Theta^i$, and that the infinitesimal
gauge transformations are generated by $G=\int d^3x
(\Lambda\Theta+\Lambda_i\Theta^i)$. The effect of gauge
transformations on the dynamical fields was obtained in
(\ref{1})-(\ref{3}). The constraints are reducible ($\partial_i
\Theta^i= \Theta$), so there is still  a residual gauge invariance
($\delta\Lambda_i=\partial_i\lambda$, $\delta\Lambda=\lambda$)
that manifests in $G$ itself. This residual invariance can be
dealt with by asking $\Lambda_i$ to be transverse
($\partial^i\Lambda_i=0)$, and by absorbing the contribution due
to the longitudinal part of $\Theta_i$ in in $\Lambda$. We can now
proceed with the generator $\overline{G}=\int d^3x
(\Lambda\Theta+{\Lambda^T}_i{\Theta^T}^i)$, whose  effect on the
dynamical fields will be analogous to the former one, with
${\Lambda^T}_i$ instead of $\Lambda_i$ in (\ref{1})-(\ref{3}).
Hence we have two points of view about the issue of gauge
invariance. In one of them we have a reducible set of constraints,
with a residual gauge invariance. In the other one, we can deal
with an irreducible set of constraints with no residual gauge
invariance. The difference between both will appear, as we shall
see, when we try to see how the multipliers change under gauge
transformations.

In order to get a complete scheme of the gauge transformations we
go to the extended action $S_E \sim \int dx(p\dot{q}-{\cal{H}})$
\cite{henneaux}, consider the gauge transformation
(\ref{1})-(\ref{3}) with time dependent gauge parameters, and
demand $S_E$ to be gauge invariant. Then, in the framework of  the
first point of view we get
\begin{eqnarray}
\delta S_E&=& \int d^4x\left[\frac{d}{dt}\left(\Lambda(x) J^0(x) +
\Lambda_i(x) J^{0i}(x)\right)\right.+\nonumber \\
&&\qquad+\left.\left(\dot{\Lambda}-\delta
A_0\right)\Theta+\left(\dot{\Lambda_i}-\delta
B_{0i}\right)\Theta^i\right],
\end{eqnarray}
so the multipliers $A_0$ and $B_{0i}$ transform as $\delta
A_0=\dot{\Lambda}$ and $\delta B_{0i}=\dot{\Lambda}_i$, provided
that the gauge parameters satisfy
\begin{equation}
\int d^3x \left(\Lambda(x) J^0(x) + \Lambda_i(x) J^{0i}(x)\right)
{\mid}^{t_{f}}_{t_{i}}=0.
\end{equation}
At this point the reducibility of the constraints plays now a role
allowing an additional invariance on the multipliers: we can add
to $B_{0i}$ the gradient of and scalar function ($\delta B_{0i}=
-\partial_i\tilde{\Lambda}$) and simultaneously transform $A_0$
($\delta A_0=-\tilde{\Lambda}$), leaving $S_E$ invariant. Taking
all this into account we conclude that the action is invariant under
the gauge transformations
\begin{equation}
\delta A_{\mu}=\partial_{\mu}\Lambda - \Lambda_{\mu} \qquad ,
\qquad \delta
B_{\mu\nu}=\partial_{\mu}\Lambda_{\nu}-\partial_{\nu}\Lambda_{\mu},
\end{equation}
where $\Lambda_{\mu}=(\tilde{\Lambda},\Lambda_i)$.

Regarding the other point of view, we also go to the extended action with the
modified gauge transformations (with $\Lambda_i$ replaced by
$\Lambda^T_i$), which leads to
\begin{eqnarray}
\delta S_E&=& \int d^4x\left[\frac{d}{dt}\left(\Lambda(x) J^0(x) +
\Lambda^T_i(x) J^{0i}(x)\right)\right.+\nonumber \\
&&\qquad+\left.\left(\dot{\Lambda}+\delta B^L-\delta
A_0\right)\Theta+\left(\dot{\Lambda}^T_i-\delta
B^T_{0i}\right){\Theta^T}^i\right],
\end{eqnarray}
where we have made the decomposition $\delta B_{0i}= \delta
B^T_{0i}+\partial_i \delta B^L$. So we obtain the transformations
$\delta A_0=\dot{\Lambda}-\delta B^L$ and $\delta
B^T_{0i}=\dot{\Lambda}^T_i$. We note that a transverse vector
$V^T_{\mu}$ ($\partial^{\mu}V^T_{\mu}=0$) can be written as
$V^T_{\mu}=((-\Delta)V, V^T_i-\partial_i\dot{V})$ with $\partial^i
V^T_i=0$. With this, the gauge transformations can be written as
\begin{eqnarray}
\delta A_0&=&\partial_0 \Lambda + \delta B^L=
\partial_0(\Lambda-\dot{\overline{\Lambda}})+ (\delta B^L+\ddot{\overline{\Lambda}}), \nonumber \\
\delta B_{0i}&=&\partial_0\Lambda^T_i+\partial_i \delta B^L=
\partial_0(\Lambda^T_i-\partial_i\dot{\overline{\Lambda}})+
\partial_i(\delta B^L+\ddot{\overline{\Lambda}}), \nonumber \\
\delta A_i&=&\partial_i \Lambda -
\Lambda^T_i=\partial_i(\Lambda-\dot{\overline{\Lambda}})
-(\Lambda^T_i-\partial_i\dot{\overline{\Lambda}}), \nonumber \\
\delta B_{ij}&=&\partial_i\Lambda^T_j-\partial_j\Lambda^T_i=
\partial_i(\Lambda^T_j-\partial_j\dot{\overline{\Lambda}})-
\partial_j(\Lambda^T_i-\partial_i\dot{\overline{\Lambda}}),
\end{eqnarray}
where we have introduced  $\overline{\Lambda}$ such
that $\delta B^L= -\square \overline{\Lambda}$. In this sense the
action is invariant under the gauge transformations
\begin{equation}
\delta A_{\mu}=\partial_{\mu}\lambda - \lambda^T_{\mu} \qquad ,
\qquad \delta
B_{\mu\nu}=\partial_{\mu}\lambda^T_{\nu}-\partial_{\nu}\lambda^T_{\mu},
\end{equation}
with $\lambda^T_{\mu}=((-\Delta)\overline{\Lambda},
\Lambda^T_i-\partial_i\dot{\overline{\Lambda}})$. These parameters
satisfy the condition
\begin{equation}
\int d^3x \left(\lambda(x) J^0(x) + \lambda^T_i(x)
J^{0i}(x)\right) {\mid}^{t_{f}}_{t_{i}}=0.
\end{equation}
In the two point of views we can eliminate $A_\mu$ and the
resulting action can be rewritten, using (\ref{2.27}), in terms of
only $a_{\mu\nu}$ as
\begin{eqnarray}
\mathcal{S}&=&\int dt \int d\sigma \frac{\alpha}{2}\lbrack
 (\dot{z}^{i})^{2}-(z'^{i})^{2}\rbrack+\nonumber \\
  & & +\int d^{4}x \left(\frac{1}{12g^{2}}H^{\mu\nu\lambda}(a)H_{\mu\nu\lambda}(a)-
  \frac{m^{2}}{4}a^{\mu\nu}a_{\mu\nu}+
  \frac{1}{2}J^{\mu\nu}a_{\mu\nu}\right).
\end{eqnarray}
So we are left with an action in terms of $a_{\mu\nu}$ and the
string coordinates $z^i$ with no further gauge invariance. The
field $a_{\mu\nu}$ describes a massive excitation of mass $\mid gm
\mid$.

{{\textbf{Acknowledgements}}}

This work was supported by Project $G-2001000712$ of FONACIT and
Project PG 03-6039-2005 of CDCH-UCV.

\end{document}